\documentclass[conference,onecolumn,draft]{IEEEtran}

\usepackage{amssymb}
\usepackage[cmex10]{amsmath}
\usepackage[final]{graphicx}
\usepackage{bm}
\newtheorem{fact}{Fact}
\newtheorem{theorem}{Theorem}

\newcommand{\A}{\mathbf{A}}
\newcommand{\F}{\mathbb{F}}
\newcommand{\G}{\mathbf{G}}

\newcommand{\I}{\mathcal{I}}
\newcommand{\N}{\mathcal{N}}

\renewcommand{\S}{\mathcal{S}}
\newcommand{\T}{\mathbf{T}}

\newcommand{\V}{\mathcal{V}}
\newcommand{\Z}{\mathcal{Z}}
\newcommand{\rank}[1]{\mathsf{rank}(#1)}
\newcommand{\al}{\alpha}
\newcommand{\X}[2]{\mathcal{X}^{\left(#1\right)}_{#2}}

\usepackage[draft=false]{hyperref}

\hypersetup{pdfauthor={Wael Halbawi},pdftitle={Distributed Reed-Solomon Codes for Simple Multiple Access Networks},pdfkeywords={Network Coding; Multiple Access Networks; Reed-Solomon Codes; Information Theory; Distributed Algorithms; Complexity; Finite Fields}}
\IEEEoverridecommandlockouts
\begin{document}

\title{Distributed Reed-Solomon Codes for Simple Multiple Access Networks}

\author{\IEEEauthorblockN{Wael Halbawi, Tracey Ho}
\IEEEauthorblockA{Department of Electrical Engineering\\
California Institute of Technology\\
Pasadena, California 91125\\
Email: \{whalbawi,tho\}@caltech.edu\\
\thanks{This work was partially supported by the Qatar Foundation - Research Division (supporting the  work of Wael Halbawi), NSF Grant CNS-0905615 (supporting the work of Tracey Ho and Hongyi Yao), and
a grant from the Simons Foundation (\#280107 to Iwan Duursma). The work of Hongyi Yao was done while he was at the California Institute of Technology.}
}
\and
\IEEEauthorblockN{Hongyi Yao}
\IEEEauthorblockA{Oracle Inc.\\
400 Oracle Parkway\\
Redwood City, California 94065\\
Email: yaohongyi03@gmail.com
}
\and
\IEEEauthorblockN{Iwan Duursma}
\IEEEauthorblockA{Department of Mathematics\\
University of Illinois\\
Urbana, Illinois 61801\\
Email: duursma@math.uiuc.edu}
}

\maketitle

\begin{abstract}
We consider a simple multiple access network in which a destination node receives information from multiple sources via a set of relay nodes. Each relay node has access to a subset of the sources, and is connected to the destination by a unit capacity link. We also assume that $z$ of the relay nodes are adversarial. We propose a computationally efficient distributed coding scheme and show that it achieves the full capacity region for up to three sources. Specifically, the relay nodes encode in a distributed fashion such that the overall codewords received at the destination are codewords from a single Reed-Solomon code.
\end{abstract}

\section{Introduction}
We consider a simple multiple access network in which a single destination node wishes to receive information from multiple sources via a set of relay nodes, each of which has access to a subset of the sources.  Each relay node is connected to the destination by a unit capacity link. Our objective is to design a distributed code that can correct arbitrary adversarial errors on up to $z$ links (or, equivalently, relay nodes). This problem has been considered previously by~\cite{yao11key} in the context of decentralized distribution of keys from a pool, where it was shown to be a special case of the general multiple access network error correction problem, whose capacity region was established in~\cite{dikaliotis11multiple}. It can also apply to other distributed data storage/retrieval scenarios where different nodes store different subsets of the source messages. 

In this paper, we propose a computationally efficient  coding scheme, distributed Reed-Solomon codes, for simple multiple access networks. In particular, the relay nodes encode in a distributed fashion such that the overall codewords received at the destination are codewords from a single Reed-Solomon code, which allows the destination to decode efficiently using classical single-source Reed-Solomon  decoding algorithms. This scheme obviates the need for encoding over successively larger nested finite fields at each source as in the prior construction of~\cite{dikaliotis11multiple}.
We prove that the proposed coding scheme achieves the full capacity region for such networks with up to three sources.

\subsection{Related work}
A related problem was studied in~\cite{Dau2013}, where the authors construct MDS codes with sparse generator matrices, motivated by sensor networks in which a group of distributed sensors collectively measure a set of conditions (sources). Unlike the scenario we study, in \cite{Dau2013} it is assumed that each sensor has access to all sources and can choose which ones to measure, and the issue of decoding complexity is not addressed.

Another related problem is the Coded Cooperative Data Exchange Problem considered in \cite{ElRouayheb2010}. Like our problem, each node has a subset of messages, but unlike our problem, the nodes communicate cooperatively via error-free broadcast transmissions in order to disseminate all messages to all nodes.

\section{System model and Background}
\begin{figure}
\centering
\includegraphics[scale=0.75]{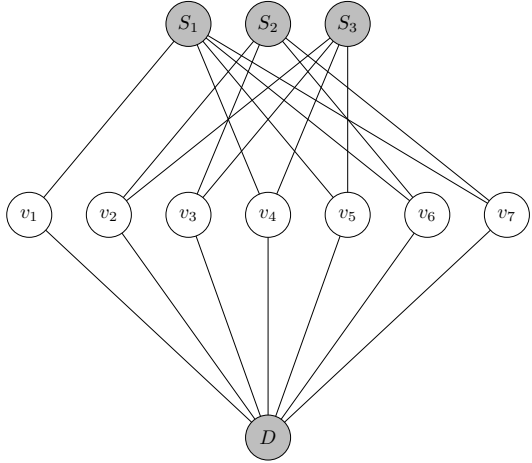}	
\caption{An example of a \textsc{sman} with 3 sources and 7 intermediate nodes.}
\label{fig:example}
\end{figure}

A Simple Multiple Access Network (\textsc{sman}) is defined as follows.  A single destination node $D$ wishes to receive information from multiple source nodes $\S = \left\{S_1,S_2, \dots, S_{|\S|} \right\}$ via a set of intermediate relay nodes $\V = \left\{v_1,\ldots,v_N\right\}$. The information rate of each source $S_i \in \S$ is denoted by $r_i$. Each relay node has access to a subset of sources, or equivalently, each source $S_i \in \S$ is connected to a subset of relay nodes by \textit{source} links of capacity $r_i$. Each relay node $v_i \in \V$ is connected to $D$ by a link of unit capacity. We refer to these links as \textit{relay} links. We wish to correct arbitrary or adversarial errors on up to $z$ relay links or equivalently nodes.  An example of a \textsc{sman} is given in Figure \ref{fig:example}.

An adjacency matrix $\A$ is associated with a \textsc{sman}, where the rows and columns represent $\S$ and $\V$, respectively, and $\A_{i,j} = 1$ if there exists a source link connecting $S_i$ to $v_j$. 

Let $\I(\S')$ denote the index set of elements in $\S'$, i.e. $\I(\S') = \left\{i:S_i \in \S'\right\}$. Also define $\I := \I(\S)$ and $r_{\I(\S')} := \sum_{i \in \I(\S')} r_i$.  The minimum cut capacity (min-cut) from $\S'$ to $D$ is denoted by $C_{\I(\S')}$, $\forall \S' \subseteq \S$. Note that $C_{\I} = N$. From~\cite{dikaliotis11multiple}, the capacity region $\mathcal{R}$ of a \textsc{sman} is given by cut set bounds for each subset of sources, i.e.~the capacity region is the set of all rate vectors $\mathbf{r} = \left(r_1,r_2,\ldots,r_{|\S|}\right)$ such that
\begin{equation}
r_{\I(\S')} \leq C_{\I(\S')} - 2z, \forall \S' \subseteq \S .\label{eqn:rate}\\
\end{equation}
Furthermore, it suffices to carry out linear network coding at internal network nodes, where each $v_i$ transmits linear combinations of received symbols over $\mathbb{F}_{q}$.

\section{Preliminaries}

To construct a distributed Reed-Solomon code for the above-described \textsc{sman} with $N$ intermediate relay nodes and $z$ adversarial nodes, we start with an ${\left[ N, k, d \right]}_q$ Reed-Solomon code where $d = 2z + 1 = N-k+1$. For the purpose of this work, we will use the definition of a Reed-Solomon Code as in~\cite{Reed1960}. This is a $k$-dimensional subspace $\mathcal{C}_\text{RS} =\left\lbrace\left[m(\alpha),m(\alpha^2),\ldots,m(\alpha^{N})\right]:\mathsf{deg\text{ }} m(x) < k \right\rbrace$, where $m(x)$ is a polynomial over $\mathbb{F}_q$ of degree $\mathsf{deg\text{ }}m(x)$, and $\alpha \in \mathbb{F}_q$ is a primitive element. The coding scheme operates over a finite field $\mathbb{F}_{q}$, where $q$ is a power of a prime $p$, such that $q \geq n + 1$. Each message vector $\mathbf{m} = [m_0,\ldots,m_{k-1}]$ is mapped to a message polynomial $m(x)=\sum_{i=0}^{k-1}m_i x^i$, which is then evaluated at $N$ distinct elements $\{\alpha,\alpha,\ldots,\alpha^{N}\}$ of $\mathbb{F}_q$. The vector of evaluations $\left[m(\alpha),m(\alpha^2),\ldots,m(\alpha^{N})\right]$ forms the corresponding codeword. This encoding operation can be described concisely using a generator matrix. The generator matrix of $\mathcal{C}_\text{RS}$ is given by $\G_\text{RS} \in \mathbb{F}_q^{k \times N}$
\begin{equation*}
\mathbf{G_\text{RS}} =
\begin{bmatrix}
1 & 1 & \cdots & 1\\
\alpha & \alpha^2 & \cdots & \alpha^{N}\\
\vdots  & \vdots & \ddots & \vdots\\
\alpha^{(k-1)}  & \alpha^{2(k-1)} & \cdots & \alpha^{N(k-1)}\\
\end{bmatrix}
\end{equation*}

For the convenience of the reader, we restate the \emph{BCH Bound} which will be used later on. For the proof, see e.g.~\cite[p.238]{McEliece}.
\begin{fact}[BCH Bound]
Let $p(x) \in \mathbb{F}[x]$ be a non-zero polynomial with $t$ (cyclically) consecutive roots, i.e. $p(\alpha^{j+1}) = \cdots = p(\alpha^{j+t})=0$. Then at least $t+1$ coefficients of $p(x)$ are non-zero.
\end{fact}
For ease of exposition, and with a slight abuse of terminology, we say that a polynomial $p(x)$ \emph{vanishes} on a set $\mathcal{P} \subseteq \{1,\ldots,N\}$ if $p(x) = \prod_{i\in\mathcal{P}}(x-\alpha^i)$.
\section{Code Construction}
As mentioned earlier, each relay node in a \textsc{sman} transmits a linear combination of its received symbols. Therefore, the overall coding operation from sources to destination can be represented by a generator matrix $\G \in \mathbb{F}_q^{r_{\I}\times N}$of a specific structure. The structure is captured by $\A$, which is used to build $\G$ as follows.  We replicate the $i^\text{th}$ row of $\A$ $r_i$ times and then replace the non-zero entries with indeterminates, whose values will be selected later on. We can write $\G$ as
\begin{equation*}
\G =
\begin{bmatrix}
\G_1 \\
\G_2 \\
\vdots\\
\G_{|\S|}
\end{bmatrix}
\end{equation*}
where the $j$th column of the submatrix $\G_i \in \mathbb{F}_q^{r_i \times N}$ is all zero if $S_i$  is not connected to $v_j$.
For a 3-source \textsc{sman}, $\G$ in generic form looks as follows:
\begin{eqnarray*}
\G &=&
\begin{bmatrix}
\G_1\\
\G_2\\
\G_3
\end{bmatrix}
=
\left[\begin{array}{c|c|c|c|c|c|c}
\bm{\times} & \mathbf{0} & \mathbf{0} & \mathbf{0} & \bm{\times} & \bm{\times} & \bm{\times} \\
\mathbf{0} & \bm{\times} & \mathbf{0} & \bm{\times} & \mathbf{0} & \bm{\times} & \bm{\times} \\
\mathbf{0} & \mathbf{0} & \bm{\times}& \bm{\times} & \bm{\times} & \mathbf{0} & \bm{\times}\\
\end{array}
\right]
\end{eqnarray*}
The symbol $\times$ represents a block of indeterminates. For example, the adjacency matrix of the \textsc{sman} in Figure \ref{fig:example} is given by
\begin{equation*}
\A = \begin{bmatrix}
1 & 0 & 0 & 1 & 1 & 1 & 1\\
0 & 1 & 1 & 0 & 0 & 1 & 1\\
0 & 1 & 1 & 1 & 1 & 0 & 0\\
\end{bmatrix}
\end{equation*}

It should be noted that the permuting the rows and/or columns of $\A$ still represents the same network. Thus, we can employ such operations when constructing a code for a certain \textsc{sman}. Now suppose $z = 1$ and $\mathbf{r} = (3,1,1)$. From $\A$, we build $\G$,

\begin{equation}
\G = \left[
\begin{array}{ccccccc}
g_{1,1} & 0 & 0 & g_{1,4} & g_{1,5} & g_{1,6} & g_{1,7} \\
g_{2,1} & 0 & 0 & g_{2,4} & g_{2,5} & g_{2,6} & g_{2,7} \\
g_{3,1} & 0 & 0 & g_{3,4} & g_{1,5} & g_{3,6} & g_{3,7} \\ \hline
0 & g_{4,2} & g_{4,3} & 0 & 0 & g_{4,6} & g_{4,7} \\ \hline
0 & g_{5,2} & g_{5,3} & g_{5,4} & g_{5,5} & 0 & 0
\end{array}
\right]
\label{eqn:mask}
\end{equation}

The indeterminates are chosen in way such that the rows of $\G$ span an $r_\I$-dimensional  subspace $\mathcal{C}$ of $\mathcal{C}_\text{RS}$. We call $\mathcal{C}$ a distributed Reed-Solomon code. For each source $i$, we can straightforwardly find a basis for the vector space of possible rows of $\G_i$  such that it spans an $r_i$-dimensional subspace of $\mathcal{C}_\text{RS}$. The only remaining question is whether  for any rate vector $\mathbf{r}$ in the capacity region it is always possible to find vectors for all $\G_i$'s so that they are collectively linearly independent. We can now describe the encoding operation. Let the  message of source $S_i$ be represented by a row vector $\mathbf{m}_i = \left[m^{(i)}_1, \ldots,m^{(i)}_{r_ i}\right]$, where $m^{(i)}_k \in \F_q$. The $j^\text{th}$ relay node encodes using the $j^\text{th}$ column $\mathbf{g}^{(j)}$ of the generator matrix $\G$, and transmits the symbol $[\mathbf{m}_1 \quad \mathbf{m}_2 \quad \dots\quad \mathbf{m}_{|\S|}]\mathbf{g}^{(j)}$. Let $\mathbf{c}$ denote the overall network codeword $\mathbf{c} = [\mathbf{m}_1 \quad \mathbf{m}_2 \quad \dots\quad \mathbf{m}_{|\S|}]\G$.
The destination node $D$ receives a corrupted version of $\mathbf{c}$, denoted by $\mathbf{y} = \mathbf{c} + \mathbf{e}$, where $\mathbf{e}$ is $z$-sparse, and $y_i$ is the symbol received by $D$ through the $i^\text{th}$ relay link. 
The following theorem  establishes that this is indeed the case for up to three sources.
\begin{theorem}For any rate vector $\mathbf{r \in \mathcal{R}}$ in the capacity region of a three-source \textsc{sman}, we can construct a distributed Reed-Solomon code.
\end{theorem}
The proof is constructive, i.e. for a given $\textsc{sman}$ along with $z$, $\mathbf{r}$ and $\A$, we show how to find a transformation matrix $\T \in \F_q^{r_\I \times k}$ such that $\G = \T\G_\text{RS}$. We can also partition $\T$ such that $\G_i = \T_i\G_\text{RS}$:

\begin{equation*}
\T =
\begin{bmatrix}
\T_1 \\
\T_2 \\
\vdots\\
\T_{|\S|}
\end{bmatrix}
\end{equation*}
We prove that such a construction is possible by considering four possible cases. Any \textsc{sman} with $\mathbf{r} \in \mathcal{R}$ falls precisely under one of these cases. For the first three cases, we show that it is always possible to set the indeterminates in a way such that $r_i$ columns in $\G_i$ form an upper triangular matrix, guaranteeing $\rank{\G} = r_i$. Given the structure of $\G_\text{RS}$, we can exactly describe $\T$ by resorting to a polynomial representation of vectors. Effectively, we transform $\G_\text{RS}$ into a matrix in row echelon form (up to a permutation of the columns). The fourth case relies on a different strategy, which is treated in a self-contained fashion. We now introduce some needed notation. For all $\S' \subseteq \S$, let $\N_{\I(\S')} \in \I(\V)$ denote the set of column indices corresponding to intermediate nodes connected to all $S_i \in \S'$ simultaneously, but not to any other source. We say $\N_{\I(\S')}$ \textit{represents} the columns indexed by its elements.  Let $n_{\I(\S')} := |\N_{\I(\S')}|$. For the network in Figure \ref{fig:example}, $\N_1 = \left\{1\right\}$, and  $\N_{1,2} = \left\{6,7\right\}$. Note that the sets $\N_{\S'}$ partition $\I(\V)$. Let $\Z_i$ be the set of indices of the columns corresponding to the relay nodes that are not connected to $S_i$. For example, $\Z_2 = \{1,4,5\}$. We will say that $\mathcal{X}\subseteq\N_{\I(\S')}$ is contained in the set of roots of a polynomial $p(x)$ if $p(\alpha^i) = 0$ for all $i \in \mathcal{X}$. We now prove the main theorem by considering each of the following four cases.

\subsection*{Case 1}
\begin{IEEEeqnarray}{rCl}
r_1 & \leq & n_1 \label{eqn:case1_r1} \\
r_2 & \leq & n_2 + n_{1,2} \label{eqn:case1_r2}
\end{IEEEeqnarray}
Without loss of generality, we assume

\begin{IEEEeqnarray*}{lCl}
\N_1 &=& \{1,\ldots,n_1\}\\
\N_2 &=&\{n_1+1,\ldots,n_1+n_2\}\\
\N_{1,2} &=& \{n_1+n_2+1,\ldots, n_1+n_2+n_{1,2}\}
\end{IEEEeqnarray*}

Given the constraint on $r_1$ in \eqref{eqn:case1_r1}, we can select $r_1$ columns represented by a subset of $\N_1$ in $\G_1$ and set as the identity matrix (or any other diagonal matrix). Similarly by \eqref{eqn:case1_r2}, a collection of $r_2$ columns represented by a subset of $\N_2\cup \N_{1,2}$ in $\G_2$ is set as a diagonal matrix. We now move to on to $\G_3$ and select any $r_3$  columns to set as a diagonal matrix. In essence, the indeterminates of $\G$ are chosen such that it is in row echelon form (up to a permutation of the columns). To show that such matrix can constructed from $\G_\text{RS}$, we define the rows of $\T$ as polynomials that vanish on appropriate sets and have the appropriate degree. Let $t^{(1)}_j(x)$ be a polynomial that vanishes on $\Z_1\cup\{1,\ldots,r_1\}\setminus j$. The $j^\text{th}$ row of $\T_1$ is the vector of coefficients of $t^{(1)}_j(x)$ along with extra zeros so that it is composed of $k$ entries. Now, let $t^{(2)}_j(x)$ be a polynomial that vanishes on $\Z_2\cup\{n_1+1,\ldots,n_1+r_2\}\setminus \left(n_1+j\right)$. $\T_2$ is built in the same manner as $\T_1$. To build $\T_3$, choose $t^{(3)}_j(x)$ such that it vanishes on $\Z_3\cup\{n_1 + n_2 + n_{1,2}+1, \ldots,n_1 + n_2 + n_{1,2}+r_3\}\setminus \left(n_1+n_2+n_{1,2}+j\right)$. Using this method, we transform $\G_\text{RS}$ into $\G$, which is in row echelon form and has no all-zero rows. The cut-set bounds \eqref{eqn:rate} along with the number of roots of $t_j^{(i)}(x)$ imply  $\mathsf{deg\text{ }}t_j^{(i)}(x)\leq k-1$. 
To see this, consider $t_j^{(1)}{(x)}$ first, which has $|\mathcal{Z}_1|+r_1-1$ roots. Since $r_1 \leq C_1-2z$ and $|\mathcal{Z}_1|=N-C_1$, we have $|\mathcal{Z}_1|+r_1-1 \leq N-2z-1 =k-1$. The same argument justifies the claim for $t_j^{(2)}(x)$ and $t_j^{(3)}(x)$. Thus, an appropriate $\T$ can always be found and  $\rank{\G} = R_\I$, as required.

\subsection*{Case 2}
\begin{IEEEeqnarray*}{rCl}
r_1 & \leq & n_1\\
r_2 & > & n_2 + n_{1,2}
\end{IEEEeqnarray*}

We make the same assumptions on $\N_1$, $\N_2$ and $\N_{1,2}$. Furthermore,
\begin{IEEEeqnarray*}{lCl}
\N_{2,3} &=& \{n_1+n_2+n_{1,2}+1,\ldots, n_1+n_2+n_{1,2}+n_{2,3}\}\\
\N_{1,2,3} &=& \{n_1+n_2+n_{1,2}+n_{2,3}+1,\ldots, n_1+n_2+n_{1,2}+n_{2,3}+n_{1,2,3}\}
\end{IEEEeqnarray*}

We define $t^{(1)}_j(x)$ and $t^{(2)}_j(x)$ as in Case 1.  Since $r_2 > n_2 + n_{1,2}$, $\T_2$ will affect, in addition to $\N_2 \cup \N_{1,2}$ columns represented by $\X{2}{} = \X{2}{2,3} \cup \X{2}{1,2,3}$, where $\X{2}{2,3} \subseteq \N_{2,3}$ and $\X{2}{1,2,3} \subseteq \N_{1,2,3}$, and $x_2 :=|\X{2}{}| = r_2-(n_2+n_{1,2})$. In other words, $\X{2}{2,3}$ is the subset of $\N_{2,3}$ contained in $\cup_j{\cal{L}}_j$, where ${\cal{L}}_j$ is the set of roots of $t_j^{(2)}(x)$. In order to have $\G$ in row echelon form, we need to modify $t^{(3)}_j(x)$ such that $\T_3$ also sets the indeterminates in the columns represented by $\X{2}{}$ in $\G_3$ to zero. Namely, $t^{(3)}_j(x)$ vanishes on $\X{2}{} \cup \Z_3 \cup \{n_1+n_2+n_{1,2}+x_2+1, \ldots,n_1+n_2+n_{1,2}+x_2+r_3\}\setminus n_1+n_2+n_{1,2}+x_2+j $. Since $t_j^{(1)}(x)$ and  $t_j^{(2)}(x)$ are chosen as in Case 1, their degrees are at most $k-1$ as required. For $t_j^{(3)}(x)$, the  following relations give the required results:

\begin{IEEEeqnarray*}{lCl}
\mathsf{deg\text{ }}t_j^{(3)}(x) &=& |\mathcal{Z}_3| + x_2 + r_3 - 1\\
&=& N- C_3 + r_2-n_2-n_{1,2} + r_3 -1\\
&=&n_1+r_2+r_3-1\\
&\leq& n_1 + C_{2,3} - 2z - 1\\
&=&N-2z-1\\
&=&k-1
\end{IEEEeqnarray*}

\subsection*{Case 3}
\begin{IEEEeqnarray}{rCl}
r_1 & > & n_1\\
r_2 & > & n_2 + n_{1,2} \label{eqn:case3_r2}
\end{IEEEeqnarray}

Assume that the elements of the index set $\I(\V)$ are ordered as follows: $\N_1,\N_{1,3},\N_2,\N_{1,2},\N_{2,3},\N_3,\N_{1,2,3}$. Given this ordering, the set $\N_{1,2}$ will be used when constructing $\G_2$. Furthemore, the columns represented by $\N_{1,3}$ are exhausted in preference to $\N_{1,2,3}$ when constructing $\T_1$. In other words, the set of roots of $t_j^{(1)}(x)$ will contain (a subset of) $\N_{1,2,3}$ only if $r_1-(n_1+n_{1,3})>0$. For $\T_2$, a similar reasoning holds by preferring $\N_{2,3}$ over $\N_{1,2,3}$. Our intuition for such ordering is that it utilizes already present zeros in $\G$ to contribute to $\rank{\G}$.
As before, $t_j^{(1)}(x)$ vanishes on $\Z_1\cup \{i_1,\ldots,i_{r_1}\}\setminus i_j$, where  $i_l$ is the $l^\text{th}$ element in the ordered set $\I(\V)\setminus \{\Z_1 \cup \N_{1,2}\}$. Note that \eqref{eqn:case3_r2} implies that the set of roots of $t_j^{(2)}(x)$ contains $\N_{1,2}$. Similarly, $t_j^{(2)}(x)$ vanishes on $\X{1}{1,2,3}\cup\Z_2\cup \{i_1,\ldots,i_{r_2}\}\setminus i_j$, where $i_l$ is the $l^\text{th}$ element in the ordered set $\bar{\Z_2}\setminus \X{1}{1,2,3}$. Lastly, $t_j^{(3)}(x)$ vanishes on $\X{1}{1,2,3}\cup \X{2}{1,2,3} \cup \X{1}{1,3}\cup\X{2}{2,3}\cup\Z_3\cup \{i_1,\ldots,i_{r_3}\}\setminus i_j$ and $i_l$ is the $l^\text{th}$ element in the ordered set $\bar{\Z_3}\setminus (\X{1}{1,2,3}\cup \X{2}{1,2,3} \cup \X{1}{1,3}\cup\X{2}{2,3})$. Before validating our choice of $t_j^{(i)}(x)$, we characterize the sizes of $\X{1}{1,3},\X{2}{2,3},\X{1}{1,2,3},\X{2}{1,2,3}$:
\begin{IEEEeqnarray*}{rCl}
\left|\X{1}{1,3}\right| &=& \mathsf{min}(n_{1,3},r_1-n_1)\\
|\X{2}{2,3}| &=& \mathsf{min}(n_{2,3},r_2-(n_2+n_{1,2}))\\
|\X{1}{1,2,3}| &=& r_1-n_1 - |\X{1}{1,3}|\\
|\X{2}{1,2,3}| &=& r_2-(n_2+n_{1,2})- |\X{2}{2,3}|
\end{IEEEeqnarray*}

We now bound the degree of $t_j^{(i)}(x)$. Using the same argument as in Case 1, $\mathsf{deg\text{ }}t_j^{(1)}(x) \leq k-1$. Now we consider $t_j^{(2)}(x)$. Assume $\left|\X{1}{1,3}\right|= n_{1,3}$:

\begin{IEEEeqnarray*}{rCl}
\mathsf{deg\text{ }}t_j^{(2)}(x) &=& |\X{1}{1,2,3}| + |\Z_2|+r_2-1\\
&=& r_1-n_1-n_{1,3}+N-C_2+r_2-1\\
&=&r_1+r_2+n_3-1\\
&\leq&C_{1,2}-2z+n_3-1\\
&=&k-1
\end{IEEEeqnarray*}
The same approach holds when justifying the claim for $t_j^{(3)}(x)$. What remains to show is that $|\X{1}{1,2,3}|+|\X{2}{1,2,3}|\leq n_{1,2,3}$. Assume $|\X{1}{1,2,3}| = r_1-n_1 -n_{1,3}$, $|\X{2}{1,2,3}| = r_2-(n_2 + n_{1,2}) - n_{2,3}$. This is justified since columns represented by elements in $\N_{1,2,3}$ are used in the construction of $\G_1$ and $\G_2$ only if these assumptions hold. By assumption on $\mathbf{r}$, $r_1 + r_2 \leq C_{1,2} - 2z \leq C_{1,2} = n_1 + n_2 + n_{1,2} + n_{1,3} + n_{2,3} + n_{1,2,3}$. Rearranging this to $r_1-n_1 -n_{1,3} + r_2-(n_2 + n_{1,2}) - n_{2,3} \leq n_{1,2,3}$ and noticing that the left hand side of the inequality is equal to $|\X{1}{1,2,3}|+|\X{2}{1,2,3}|$ yields the result.
\subsection*{Case 4}
\begin{IEEEeqnarray*}{rCl}
r_1 & > & n_1\\
r_2 & \leq & n_2 + n_{1,2}
\end{IEEEeqnarray*}

This case is approached differently. First, we permute the $\G_i$'s to recast the problem such that it falls under Case 1, 2 or 3. If this is possible, then we basically construct the code as earlier. Otherwise, for all $i \neq j$,
\begin{IEEEeqnarray}{rCl}
r_i &>& n_i\\
r_i &\leq& n_i + n_{i,j}  \label{eqn:case4}
\end{IEEEeqnarray}
We will assume that the columns of $\G$ are ordered in the following manner.
\begin{equation*}
\G =
\begin{bmatrix}
\G_1\\
\G_2\\
\G_3
\end{bmatrix}
=
\left[\begin{array}{ccc|ccc|c}
\bm{\times} & \mathbf{0} & \mathbf{0} &  \bm{\times} & \bm{\times} & \mathbf{0} &\bm{\times} \\
\mathbf{0} & \bm{\times} & \mathbf{0} & \bm{\times} & \mathbf{0} & \bm{\times} & \bm{\times} \\
\mathbf{0} & \mathbf{0} & \bm{\times} & \mathbf{0} & \bm{\times} & \bm{\times}  & \bm{\times}\\
\end{array}
\right]
\end{equation*}
For this case, we place an identity of size $n_i$ in the submatrix that corresponds to the columns represented by $\N_i$ and the block $\G_i$. We permute the rows of $\G$ to obtain the following:
\begin{equation}
\G =
\left[\begin{array}{ccc|ccc|c}
\mathbf{I}& \mathbf{0} & \mathbf{0} & \bm{\times} & \bm{\times} & \mathbf{0}&  \bm{\times} \\
\mathbf{0} & \mathbf{I}& \mathbf{0} & \bm{\times} & \mathbf{0} & \bm{\times} & \bm{\times} \\
\mathbf{0} & \mathbf{0}& \mathbf{I} & \mathbf{0} & \bm{\times} & \bm{\times} & \bm{\times}\\\hline
\mathbf{0}& \mathbf{0} & \mathbf{0}  & \bm{\times} & \bm{\times} & \mathbf{0}& \bm{\times} \\
\mathbf{0} & \mathbf{0}& \mathbf{0} & \bm{\times} & \mathbf{0} & \bm{\times} & \bm{\times} \\
\mathbf{0} & \mathbf{0} & \mathbf{0}& \mathbf{0} & \bm{\times} & \bm{\times}  & \bm{\times}\\
\end{array}
\right]\label{eqn:mask_case4}
\end{equation}
The blocks of columns of $\G$ have sizes $n_{1}, n_{2}, n_{3}, n_{1,2}, n_{1,3}, n_{2,3}, n_{1,2,3}$, while the blocks of rows have sizes $n_1, n_2, n_3, r_1-n_1, r_2 - n_2, r_3 - n_3$.

We will construct a matrix $\T$ (with dimensions $R_\I \times k$), partitioned into two row-blocks, that will transform the generator matrix of a  $\left[N, k, d\right]$ RS code $\G_{\text{RS}}$ to one whose mask is given by $\G$ in \eqref{eqn:mask_case4}.

\begin{equation}
\G = \T\G_\text{RS}  =
\begin{bmatrix}
\mathbf{S} \\ \mathbf{V}
\end{bmatrix}
\G_\text{RS}
=
\left[
\begin{array}{c|c}
\mathbf{I} & \mathbf{\hat{G}_1} \\\hline
\mathbf{0} & \mathbf{\hat{G}_2} \\
\end{array}
\right]
=\begin{bmatrix}
\mathbf{X}\\ \hat{\G}\\
\end{bmatrix} \label{eqn:G_hat}
\end{equation}
 Note that $\mathbf{S}$ has $\bar{n}:=\sum n_i$ rows and $\mathbf{V}$ has $R':=\sum r'_i$ rows, where $r_i' = r_i - n_i$.  Constructing $\mathbf{S}$ is straightforward and we only focus on $\mathbf{V}$.

\subsubsection*{Construction of $\mathbf{V}$}
As before, the construction boils down to finding polynomials with the appropriate roots, i.e. roots corresponding to the zeros in $\hat{\G}$. Each of these polynomials will have $N-(1+2z)$ roots. Using the coefficients of these polynomials as the rows of $\mathbf{V}$ will provide the required result. Let $c(x) = \prod_{i=1}^{\bar{n}}(x-\alpha^i)$. This polynomial will produce the all-zero columns of $\hat{\G}$, namely the columns represented by $\N_1 \cup \N_2 \cup \N_3$. Now consider the polynomial $p(x) = \prod_{i \in \mathcal{P}}(x-\alpha^i)$ which vanishes on $\mathcal{P} = \{n_{1,2} + n_{1,3} + \bar{n} + 1,n_{1,2} + n_{1,3}+ \bar{n}+2,\ldots, n_{1,2} + n_{1,3}  + k - 1\}$, and form the polynomial $v(x) = c(x)p(x)$. Defining the row vector $[v(\alpha^i)]$, $i = 1,\ldots,N$ yields the first row of $\hat{\G}$ in \eqref{eqn:G_hat}.

To proceed with the construction,we define the following sets:

\begin{eqnarray*}
\mathcal{J}_1 &=& \left\lbrace 0,1,\ldots, r'_1 -1 \right\rbrace\\
\mathcal{J}_2 &=& \left\lbrace n_{1,3},n_{1,3}+1,\ldots, n_{1,3}+r'_2-1\right\rbrace\\
\mathcal{J}_3 &=& \left\lbrace n_{1,3}+n_{1,2},n_{1,3}+n_{1,2}+1,\ldots, n_{1,3}+n_{1,2}+r'_3-1\right\rbrace
\end{eqnarray*}

We partition $\mathbf{V}$ as follows:
\begin{equation*}
\mathbf{V} =
\begin{bmatrix}
\mathbf{V}_1\\
\mathbf{V}_2\\
\mathbf{V}_3
\end{bmatrix}
\end{equation*}
The $j^\text{th}$ row of $\mathbf{V}_i$ (with dimensions $r'_i \times k$) corresponds to the coefficients of $v^{(i)}_j(x) = c(x)p(\alpha^{j}x)$, where $j \in \mathcal{J}_i$. Using this method, $c(x)$ still produces the zeros required for the all zero block. The remaining zeros in each row are produced by $p(\alpha^{j}x)$, which basically shifts the location of the roots of $p(x)$ by $j$ positions to the left appropriately. Since $p(x)$ has $ t=k-\bar{n}-1$ roots, a row in $\hat{\G}_2$ with a requirement of, say, $n_{2,3}$ zeros will have an excess of $t-n_{2,3}$ zeros. Nonetheless, the weight of every row of $\hat{\G}$ is still at least $1+2z$. Now, we proceed to show that $\mathbf{V}$ is full rank.

First, we need that the $\mathcal{J}_1,\mathcal{J}_2,\mathcal{J}_3$ sets are pairwise disjoint. Note the elements in each $\mathcal{J}_i$ are increasing. By the constraints \eqref{eqn:case4}, $r_1 \leq n_1 + n_{1,3}$. Thus $r'_1 - 1 < n_{1,3}$ and $\mathcal{J}_1 \cap \mathcal{J}_2 = \emptyset$. A similar argument implies $\mathcal{J}_2 \cap \mathcal{J}_3 = \emptyset$.

Now, we show that the polynomials $v^{(i)}_j(x)$ are linearly independent. Note that this is true if and only if the polynomials $p(\alpha^jx)$ are linearly independent. Therefore, we focus on the latter.

\section{Rank of \texorpdfstring{$\mathbf{V}$}{V}}
Write $p(x) = \sum_{l=0}^{t} p_l x^l$ and $p(\alpha^{j}x) = \sum_{l=0}^{t} p_l \alpha^{jl}x^l$. Consider the matrix $\mathbf{P}$ formed by the coefficients of $p(\alpha^{j}x)$, $j\in \cup_i\mathcal{J}_i$

\begin{equation*}
\mathbf{P} =
\begin{bmatrix}
p_0 & p_1\alpha^{j_1} & \ldots & p_{t}\alpha^{j_1t}\\
p_0 & p_1\alpha^{j_2} & \ldots & p_{t}\alpha^{j_2t}\\
\vdots & \vdots & \ddots & \vdots\\
p_0 & p_1\alpha^{j_{R'}} & \ldots & p_{t}\alpha^{j_{R'}t}\\
\end{bmatrix}
\end{equation*}

The matrix $\mathbf{P}$ (dimensions $R' \times t+1$) is never a tall matrix since by the rate region $R' \leq t+1$. Consider the matrix $\hat{\mathbf{P}}$ which is formed from the first $R'$ columns of $\mathbf{P}$. Writing out the determinant of $\hat{\mathbf{P}}$ yields
\begin{IEEEeqnarray*}{rCl}
\mathsf{det}(\hat{\mathbf{P}}) &=&
\begin{vmatrix}
p_0 & p_1\alpha^{j_1} & \ldots & p_{R-1}\alpha^{j_1(R'-1)}\\
p_0 & p_1\alpha^{j_2} & \ldots & p_{R-1}\alpha^{j_2(R'-1)}\\
\vdots & \vdots & \ddots & \vdots\\
p_0 & p_1\alpha^{j_{R'}} & \ldots & p_{R'-1}\alpha^{j_{R'}(R'-1)}\\
\end{vmatrix}\\
&=&
\prod_{i = 0}^{R'-1}p_i
\begin{vmatrix}
1 & \alpha^{j_1} & \ldots & \alpha^{j_1(R'-1)}\\
1 & \alpha^{j_2} & \ldots & \alpha^{j_2(R'-1)}\\
\vdots & \vdots & \ddots & \vdots\\
1 & \alpha^{j_{R'}} & \ldots & \alpha^{j_{R'}(R'-1)}\\
\end{vmatrix}
\end{IEEEeqnarray*}

Using the BCH bound, we establish that all $p_i$'s are nonzero. Therefore $\mathsf{det}(\hat{\mathbf{P}})$ is equal to the determinant of the Vandermonde matrix with defining set $\{\alpha^{j_1}, \ldots, \alpha^{j_{R'}}\}$, multiplied by a non-zero scalar. As it was established earlier, the elements $\{\alpha^{j_1}, \ldots, \alpha^{j_{R'}}\}$ are all distinct in $\mathbb{F}_q$.

Therefore, $\mathbf{P}$ is full rank implying that the polynomials $v_j^{(i)}(x)$ are linearly independent and so $\rank{\hat{\G}} = R'$. We also observe that $\rank{\G} = \rank{\mathbf{X}} + \rank{\hat{\G}} = \bar{n} + R' = R_\I$. Thus $\G$ is full rank.
\section{Decoding Details}
In essence, we have used a subcode of a RS code to correct errors in a \textsc{sman}. Generally speaking, any decoding algorithm designed for \emph{evaluation-based} construction of RS-codes can be used to decode the received symbols to a valid codeword $\hat{\mathbf{c}}$. However, this doesn't imply that the we can immediately \emph{de-map} the codeword to the original message $\mathbf{m}$. Nonetheless, if we were to use a decoder that produces a valid message $\hat{\mathbf{m}}_\text{RS}$ such that $\hat{\mathbf{c}}=\hat{\mathbf{m}}_\text{RS}\G_\text{RS}$, then we can obtain an estimate of the transmitted message vector $\hat{\mathbf{m}}$ easily using our knowledge of $\T$.

Thus, we use the Berlekamp-Welch decoder as described by Gemmell and Sudan in \cite{Gemmell1992}. Assuming at most $z$ errors occur, we recover the unique $\mathbf{m}_\text{RS}$ such that $\mathbf{c} = \mathbf{m}\G = \mathbf{m}\T\G_\text{RS}= \mathbf{m}_\text{RS}\G_\text{RS}$. Since $\T$ has full row rank, it follows that $\mathbf{m}_\text{RS} =\mathbf{m}\T$. Next let $\tilde{\T}$ be a matrix of $R_\I$ columns of $\T$ such that it is invertible\footnote{If the problem instance falls under Case 4, then we know that the first $R_\I$ columns are linearly independent. Otherwise, $\T$ can be reduced to row echelon form using Gaussian elimination, and then we select the pivot columns.}. We can now recover $\mathbf{m}$ since $\tilde{\mathbf{m}}_\text{RS}\tilde{\T}^{-1} = \mathbf{m}$, where $\tilde{\mathbf{m}}_\text{RS}$ is a subvector of $\mathbf{m}_\text{RS}$ with length $R_\I$, corresponding to the columns of $\T$ selected in $\tilde{\T}$.
\section{Example}
In this section, we show how to construct a DRS code for the \textsc{sman} in Figure \ref{fig:example}. Assume $\mathbf{r} = (3,1,1)$ and $z=1$. From here, it follows that the construction of Case 4 should be used. The constituent code is a $[7,5,3]$ RS code over $\mathbb{F}_8$ with primitive polynomial $x^3 + x + 1$. The poynomials and sets of interest are tabulated below:

\begin{IEEEeqnarray*}{rCl}
c(x) &=& (x-\alpha)\\
p(x) &=& (x-\alpha^6)(x-\alpha^7)(x-\alpha^8)\\
\mathcal{J}_1 &=& \{0,1\}\\
\mathcal{J}_2 &=& \{2\}\\
\mathcal{J}_3 &=& \{4\}\\
\end{IEEEeqnarray*}

Next we can evaluate $v_j^{(i)}(x)$ for $j\in \mathcal{J}_i$ and $i=1,2,3$ to obtain

\begin{equation*}
\mathbf{V} =
\begin{bmatrix}
\al & \al^4 & 1 & \al^2 & 1\\
\al & \al^2 & \al^4 & \al^3 & \al^3 \\
\al & 0 & 1 & \al^3 & \al^6\\
\al &\al^6  & \al^5 & \al & \al^5
\end{bmatrix}
\end{equation*}

The polynomial $s(x)$ corresponding to $\mathbf{S}$ is $s(x) = \frac{(x-\alpha^6)(x-\alpha^7)}{(\al-\alpha^6)(\al -\alpha^7)}$. The scaling factor forces $s(\al) = 1$. Finally, we have the transformation matrix $\T$ and the corresponding $\G$:
\begin{equation*}
\mathbf{\T} =
\left[
\begin{array}{ccccc}
\al^5 & \al & \al^6 & 0 & 0\\\hline
\al & \al^4 & 1 & \al^2 & 1\\
\al & \al^2 & \al^4 & \al^3 & \al^3 \\
\al & 0 & 1 & \al^3 & \al^6\\
\al &\al^6  & \al^5 & \al & \al^5
\end{array}\right]
\end{equation*}

\begin{equation*}
\mathbf{\G} =
\left[
\begin{array}{ccccccc}
1 & \al^5 & \al^4 & 1 & \al^4 & \mathbf{0} & \mathbf{0} \\
 0 & 1 & \al^5 & \al^5 & \al^3 & \mathbf{0} & \mathbf{0} \\
0 & \al^{2} & \al^3 & \al^6 & 0 & \mathbf{0} & \mathbf{0} \\ \hline
\mathbf{0} & 1 & \al^4 & \mathbf{0} & \mathbf{0} & 0 & \al^6 \\ \hline
\mathbf{0} & \mathbf{0} & \mathbf{0} & 0 & \al^{2} & \al^3 & \al^6
\end{array}
\right]
\end{equation*}

The required zeros are in boldface. Note that $\G$ has the same form of the following mask, which is a valid permutation of the generic mask in \eqref{eqn:mask}.

\begin{equation*}
\G = \left[
\begin{array}{ccccccc}
g_{1,1} & g_{1,2} & g_{1,3} & g_{1,4} & g_{1,5}& \mathbf{0} & \mathbf{0}  \\
g_{2,1} & g_{2,2} & g_{2,3} & g_{2,4} & g_{2,5}& \mathbf{0} & \mathbf{0}  \\
g_{3,1} & g_{3,2} & g_{1,3} & g_{3,4} & g_{3,5}& \mathbf{0} & \mathbf{0}  \\ \hline
\mathbf{0} & g_{4,2} & g_{4,3} & \mathbf{0} & \mathbf{0} & g_{4,6} & g_{4,7} \\ \hline
\mathbf{0} & \mathbf{0} & \mathbf{0} & g_{5,4} & g_{5,5} & g_{5,6} & g_{5,7}
\end{array}
\right]
\end{equation*}

\section{Conclusion}
We have proposed a distributed Reed-Solomon coding scheme for simple multiple access networks, with much lower decoding complexity compared to existing constructions for the general multiple access network error correction problem. The field size scales linearly with the length of the code as opposed to exponentially with the number of sources. We show that it achieves the full capacity region for up to three sources. It remains as further work to determine whether it can achieve the full capacity region for networks with more than three sources; our proof for three sources does not extend straightforwardly since the method of classifying the problem depending on the rate vector results in an exponentially growing number of cases.
\bibliographystyle{IEEEtran}
\bibliography{IEEEabrv,library,NWC-abbr}

\end{document}